\documentstyle{article}
\renewcommand{\theequation}{\thesection.\arabic{equation}}
\newcommand\beq{\begin{equation}}
\newcommand\eeq{\end{equation}}
\newcommand{\eqref}[1]{(\ref{#1})}

\def\beqa{\begin{eqnarray}}
\def\eeqa{\end{eqnarray}}
\def\ba{\begin{array}}
\def\ea{\end{array}}
\def\r{\rangle}
\def\l{\langle}
\def\a{\alpha}
\def\b{\beta}

\def\d{\delta}
\def\g{\gamma}
\def\det{\mbox{det}}

\def\s{\sigma}
\def\sl{\sum\limits}
\def\pl{\prod\limits}
\def\lt({\left(}
\def\rt){\right)}
\def\pd #1{\frac{\partial}{\partial #1}}

\newcommand{\bl}[1]{\makebox[#1em]{}}
\begin{document}
\author{A.G. Izergin\thanks{e-mail: izergin@pdmi.ras.ru},\\{\small
St. Petersburg Department of the Steklov Mathematical Institute
(POMI),}\\ {\small Fontanka 27, St. Petersburg 191011, Russia}\\ \\ V.S.
Kapitonov\thanks{e-mail: kapiton@tu.spb.ru},\\
{\small St. Petersburg Technological Institute,
Moskovski pr., 26 St. Petersburg 198013, Russia} \\ \\N.A.
Kitanine\thanks{e-mail:  kitanin@pdmi.ras.ru}\\ {\small St.  Petersburg
Department of the Steklov Mathematical Institute (POMI),}\\ {\small
Fontanka 27, St.  Petersburg 191011, Russia and}\\ {\small Laboratoire
de Physique Th\'eorique {\sc enslapp} ENSLyon, 46 All\'ee d'Italie
69007 Lyon, France}}

\title{Equal-time temperature correlators of the
one-dimensional Heisenberg XY chain}

\date{}

\maketitle

\abstract{
Representations as determinants of $M\times M$ dimensional matrices are
obtained for equal-time temperature correlators of the anisotropic
Heisenberg XY chain. These representations are simple deformations of
the answers for the isotropic XX0 chain.  In the thermodynamic limit,
the correlators are expressed in terms of the Fredholm determinants of
linear integral operators.}

\section*{Introduction}
The description of
the correlation functions for the models solved by means of Bethe
 ansatz is based on the representation for the correlators as the
Fredholm determinants of linear integral operators. Such
representations were obtained for the first time in \cite{l1,l2} for
the simplest two point equal-time correlators for the one-dimensional
impenetrable bosons model. Later they were generalized for the case of
time-dependent correlators for the models being the free-fermion point
of models solved by means of Bethe Ansatz (impenetrable bosons
\cite{ks} and the isotropic XX0 Heisenberg chain \cite{cikt1,cikt2})
and also for the case of finite interaction (\cite{k,kks} for the
one-dimensional Bose gas and \cite{efik} for the XXZ chain). Such
representations allow to write classical integrable equations for
correlators which can be used, in particular, to calculate the long
time and large distance asymptotics for the correlation functions
\cite{iik1,iiks2,iik2,iikv}.

 It was understood a time ago \cite{TMcC,WMcC,McCTW}
that the language of the classical differential equations is
natural for the description of the correlation functions
of quantum integrable models. The recent progress in this direction
is described in detail in the book \cite{bik} for the example
of impenetrable bosons based on the approach elaborated
in \cite{iik1,iiks1,iiks2,iiks3,iik2,iikv}. The most important
part of this approach is to consider the Fredholm determinant
of the linear operator appearing in the representation of
the correlators as a $\tau$-function for a new system of classical
integrable equations (see also \cite{jmms}). These linear operators
should be of a very special form - so called "integrable"
integral operators (see \cite{bik,iiks4}).

In this paper  we obtain some determinant representations
for equal-time temperature correlators for the anisotropic
Heisenberg XY chain both for finite lattice and in thermodynamic
limit.  This model was introduced and studied for the first
time in \cite{lsm}. Later it was investigated by many authors
(see \cite{McC,McCBA,McCPS} and references therein).

To compute the correlation functions we use a modification
of the approach proposed in \cite{ki} and based on using
the integration over the Grassman variables and the corresponding
coherent states. It is an essential point in our paper.
We don't use, however, the functional integral as in  \cite{ki}.
Using the coherent states simplifies the calculations
leading automatically to the answers in the necessary form
(see for example Appendix B where we reproduce the results
\cite{cikt1,cikt2} for the XX0 chain using our method).

 The representations for the equal-time
temperature correlators of the anisotropic XY chain
as determinants of $M\times M$ matrices obtained in this paper
being a direct generalization of the representations for
the isotropic case \cite{cikt1,cikt2,cit} differ from them
only by changing "Fermi" (or "Bose") weight in the kernels
of integral operators for a weight depending on the anisotropy
parameter.  On the other hand these representations generalize the
results for the anisotropic chain obtained in \cite{iks} for zero
temperature.  The "integrability" of the integral operators appearing
in the thermodynamic limit is evident.

 One should note that the isotropic XX0 chain is the "free-fermion
point" for the XXZ Heisenberg chain which is a model solved by means
of the standard Bethe Ansatz. The anisotropic XY chain is the
"free-fermion line" for the XYZ chain which is a model where
the usual Bethe Ansatz doesn't work. Therefore it is very
interesting from our point of view that the answers for the
equal-time correlators have the form of a simple deformation
of the answers for the isotropic case. We hope to give a
corresponding description for the time-depending correlation
functions in our next publication.

Our paper is organized as follows.

In the first section we describe the model and give the basic
facts about the diagonalization of the Hamiltonian using the Bogolyubov
transformation following the classical works \cite{lsm,McCBA}. In
section 2 we introduce the coherent states for isotropic
XX0 model and calculate the matrix elements of operators
between these states. In section 3 we describe the coherent
states for the anisotropic XY chain and consider their relations
with the coherent states of the isotropic chain. In section 4
the simplest correlator is calculated. In section 5 we calculate
the local spin correlator for the finite anisotropic lattice.
In section 6 we obtain the results in the thermodynamic limit.
In appendix A the basic facts about Grassmanian coherent states
are given. In appendix B the derivation of the results for the
isotropic case is presented.

\section{The XY Heisenberg chain}
\setcounter{equation}{0}

The Hamiltonian of the XY spin chain describing the interaction
between the nearest
neighbors spins 1/2 placed in the sites of  one-dimensional periodical
lattice in a constant magnetic field $h$ is
\beq
{\bf H}={\bf H}_0+\g{\bf H}_1-h{\bf S}^z,
\label{(1.1)}
\eeq
where
\beq
\label{(1.2)}
{\bf H}_0=-\frac 12 \sl_{m=1}^M (\s^+_m\s^-_{m+1}+\s^-_m\s^+_{m+1});
\eeq
\beq
\label{(1.3)}
{\bf H}_1=-\frac 12 \sl_{m=1}^M (\s^+_m\s^+_{m+1}+\s^-_m\s^-_{m+1});
\eeq
and the third component of the total spin is
\beq
\label{(1.4)}
{\bf S}^z=\sl_{m=1}^M\s_m^z.
\eeq

The total number of sites $M$ is supposed to be even. Pauli matrices
are defined as usual
\beq
\label{(1.5)}
[\s^\a_m,\s^\b_n]=2i\d_{mn}\epsilon^{\a\b\g}\s^\g_m (\a,\b,\g=x,y,z);
\eeq
\[\s^\pm_m=\frac 12(\s_m^x\pm i\s_m^y),\]
with the periodical boundary conditions
\beq
\label{(1.6)}
\s^\a_{M+1}=\s^\a_1.
\eeq

Due to the symmetries of the Hamiltonian the sign of the magnetic field
(as the sign of the Hamiltonian) is not essential. We will assume that
$h\ge 0$.

The Jordan-Wigner transformation
\[a_m=\exp \left\{ i\pi{\bf Q}(m-1)\right\}\s^+_m;\]
\beq
a^+_m=\s^-_m\exp \left\{ i\pi{\bf Q}(m-1)\right\},
\label{(1.7)}
\eeq
introduces the canonical fermion fields $a_m,a_m^+$ on the lattice,
\[ [a_m,a_n]_+\equiv a_m a_n + a_n a_m= 0;\]
\beq
\label{(1.8)}
[a^+_m,a^+_n]_+=0, [a_m,a^+_n]_+=\d_{mn}.
\eeq

Operator ${\bf Q}(m)$ is the operator of number of particles on the
first $m$ sites of the lattice,
\beq
\label{(1.9)}
{\bf Q}(m)=\sl_{j=1}^m q_j, \eeq where $q_m$ is the operator of number
of particles in the site $m$:  \beq \label{(1.10)} q_m=a^+_m
a_m=\s_m^-\s_m^+=\frac 12 (1-\s^z_m).  \eeq

 The operator of the
total number of particles,
\beq
\label{(1.11)}
{\bf N}={\bf Q}(M),
\eeq
commutes with the operators ${\bf H}_0$ and ${\bf S}^z$ but does not
commute with the operator ${\bf H}_1$ and thus the total Hamiltonian
${\bf H}$ does not conserve the number of  "$a$-fermions". At the same
time the operator $(-1)^{\bf N}=\exp \{\pm i\pi {\bf N}\}$,
anticommuting with the creation and annihilation operators
\beq
\label{(1.12)}
[(-1)^N,a^+_m]_+=[(-1)^N,a_m]_+=0,
\eeq
commutes with
any bilinear in $a_m,a_m^+$ operators, in particular, with the
Hamiltonian \beq \label{(1.13)} [(-1)^{\bf N},{\bf H}]=0.  \eeq

 Periodical boundary conditions \eqref{(1.6)} for the spins lead
to the following conditions for the fermions:
\beq
\label{(1.14)}
a_{M+1}=(-1)^{\bf N} a_1; \bl{1} a^+_{M+1}=a^+_1 (-1)^{\bf N}.
\eeq
Introducing projectors ${\bf P^\pm}$
\[{\bf P^\pm}=\frac 12 (1\pm (-1)^{\bf N});\]
\beq
\label{(1.15)}
({\bf P^\pm})^2={\bf P^\pm}; \bl{1} {\bf P^+}+{\bf P^-}=1; \bl{1}
{\bf P^+}{\bf P^-}={\bf P^-}{\bf P^+}=0;
\eeq
\[{\bf P^\pm}a_m=a_m{\bf P^\mp}; \bl{1} [{\bf H},{\bf P^\pm}]=0,\]
one can rewrite the Hamiltonian in the following form \cite{McCBA}
\beq
\label{(1.16)}
{\bf H}={\bf H^+}{\bf P^+}+{\bf H^-}{\bf P^-}.
\eeq
Both operators ${\bf H}^\pm$ can be rewritten formally in the same form
\beq
\label{(1.17)}
{\bf H^\pm}=\frac 12 \sl_{m=1}^M \left[ (a^+_m a_{m+1}+ a^+_{m+1} a_m)+
\g (a^+_m a^+_{m+1}+ a_{m+1} a_m)\right]+
\eeq
\[+h\sl_{m=1}^M a^+_m a_m-
\frac{hM}{2},\]
the only difference between ${\bf H^+}$ and ${\bf H^-}$ being the
boundary conditions:  \[a_{M+1}=-a_1;\bl{1} a^+_{M+1}=-a^+_1;\bl{1}
{\rm for} \bl{1}{\bf H^+},\] \beq \label{(1.18)} a_{M+1}=a_1;\bl{1}
a^+_{M+1}=a^+_1;\bl{1} {\rm for} \bl{1} {\bf H^-}.  \eeq

 Hence the Fourier transformations to the momentum representation are
different for these Hamiltonians. We
denote the sets of permitted quasimomenta $X^+$ for ${\bf H^+}$ and
$X^-$ for ${\bf H^-}$:
\beq
\label{(1.19)}
X^\pm=\left\{ p: \exp\{ ipM\}=\mp 1,\bl{1} p\in (-\pi,\pi]\right\},
\eeq
or explicitly:
\[X^+=\left\{ p_l=-\pi -\frac{\pi}{M}+
\frac{2\pi}{M}l, \bl{1}l=1,2,\dots,M\right\},\]
\beq
X^-=\left\{ p_l=-\pi -\frac{2\pi}{M}l, \bl{1}l=1,2,\dots,M\right\}.
\label{(1.20)}
\eeq
The corresponding formulae for the Fourier transformation can be
written in the following form
\[a_m=\frac{\exp \{-i\pi /4\}}{\sqrt{M}}\sl_{p\in X^\pm} a_p
\exp\left\{ i(m-1)p\right\},\]
\beq
\label{(1.21)}
a^+_m=\frac{\exp \{i\pi /4\}}{\sqrt{M}}\sl_{p\in X^\pm} a^+_p
\exp\left\{ -i(m-1)p\right\},
\eeq
(the summation is taken over $p\in X^+$ for ${\bf H^+}$ and over $p\in
X^-$ for ${\bf H^-}$) and \[a_p=\frac{\exp \{i\pi
/4\}}{\sqrt{M}}\sl_{m=1}^M a_m \exp\left\{ -i(m-1)p\right\},\] \beq
\label{(1.22)}
a^+_p=\frac{\exp \{-i\pi /4\}}{\sqrt{M}}\sl_{m=1}^M a^+_m
\exp\left\{ i(m-1)p\right\}.
\eeq
The Hamiltonians ${\bf H^\pm}$ in
the momenta representation are written as
\beq
\label{(1.23)}
{\bf H^\pm}=\sl_{p\in X^\pm}\left[\varepsilon (p)a^+_p a_p+
\frac {\Gamma (p)}{2}(a^+_p a^+_{-p}+a_{-p}a_p)\right]-\frac{Mh}{2},
\eeq
where
\beq
\label{(1.24)}
\varepsilon (p)=h-\cos p;\bl{1} \Gamma (p)=\g\sin p.
\eeq
 Diagonalization of these Hamiltonians can be done using
the  Bogolyubov transformation (different for ${\bf H^+}$ and ${\bf H^-}$)
leading to new canonical fermion operators $A_p$ and $A_p^+$
\beq
\label{(1.25)}
\ba{c}
A_p=\a (p)a_p-\b (p)a^+_{-p};\\
A^+_p=\a (p)a^+_p+\b (p)a_{-p},
\ea
\eeq
where
\beq
\a(p)=\cos \frac{\theta (p)}{2};\bl{1}\b(p)=\sin\frac{\theta (p)}{2},
\label{(1.26)}
\eeq
and the angle $\theta (p)$ is defined by relations:
\beq
\ba{c}
\cos\theta (p)=\frac{\varepsilon (p)}{E(p)};\bl{1}
\sin\theta (p)=-\frac{\Gamma (p)}{E(p)};\\
p\neq 0,\pi; \theta (p)=-\theta (-p),
\ea
\label{(1.27)}
\eeq
\beq
\label{(1.28)}
E(p)=\sqrt{\varepsilon^2(p)+\Gamma^2(p)}\geq0, \bl{1}(p\neq0,\pi).
\eeq
The momenta $p=0,\pi$ (appearing only in ${\bf H^-}$) should be treated
separately. Following \cite{McCBA} we put
\beq
\label{(1.29)}
\ba{c}
A_0=a_0;\bl{1} A^+_0=a^+_0,\\
A_\pi=a_\pi;\bl{1} A^+_\pi=a^+_\pi,
\ea
\eeq
and
\beq
\label{(1.30)}
\ba{c}
E(0)=h-1=\varepsilon (0)\bl{1} (E(0)<0 \bl{1} {\rm for} \bl{1} h<1)\\
E(\pi)=h+1=\varepsilon (\pi)>0.
\ea
\eeq

The Hamiltonians can be diagonalized as follows
\beq
\label{(1.31)}
{\bf H^\pm}=\sl_{p\in X^\pm}E(p)A^+_p A_p+E_0^\pm,
\eeq
where the "vacuum energy" is
\beq
\label{(1.32)}
E_0^\pm=-\frac 12 \sl_{p\in X^\pm}E(p),
\eeq
(to calculate $E_0^-$ one should take into account the definition (1.30)).
One should note that  for $h<1$ the value $E_0^-$ is not the ground
state energy $E_g^-$, in this case $E_g^-=E_0^-+\varepsilon (0)$.

\section{Coherent states for the XX0 chain}
\setcounter{equation}{0}
In this section we give some formulae for  matrix elements of
operators between the coherent states of the isotropic XX0 chain which
are necessary for the following calculations. The corresponding
Hamiltonians ${\bf H}_{\rm XX0}^\pm$ (see \eqref{(1.23)})
\beq
\label{(2.1)}
{\bf H^\pm_{\rm XX0}}=\sl_{p\in X^\pm} \varepsilon (p)a^+_p a_p-\frac{hM}{2}
\eeq
will be denoted simply ${\bf H^\pm}$ in this section for the
simplification.  They are diagonal already in terms of the operators
$a_p,a_p^+$ \eqref{(1.21)}. One should note, however, that the Fock
vacuum (which is the same for ${\bf H^\pm}$),
\beq
\label{(2.2)}
a_m|0\r=0, \bl{1}
\l 0|a^+_m =0 \bl{1} (m=1,2,\dots ,M),
\eeq
\[a_p|0\r=0, \bl{1} \l
0|a^+_p =0, \bl{1} (p\in X^\pm ),\bl{2}\l 0|0\r=1,\]
is the ground state
only for $h>1$.

 One introduces the coherent states (see Appendix A)  different for the
Hamiltonians ${\bf H^+}$ and ${\bf H^-}$
\beq
\label{(2.3)}
|\phi,\pm\r =\exp\left\{\sl_{q\in X^\pm} a^+_q\phi_q\right\}|0\r,
\eeq
\beq
\label{(2.4)}
\l\phi^*,\pm| =\l0|\exp\left\{\sl_{q\in X^\pm}\phi^*_q a_q\right\}.
\eeq
The parameters $\phi_q,\phi_q^*$ (Grassman algebra elements)
anticommute with other parameters and with all the operators
$a, a^+$.  The main properties of the coherent states \eqref{(2.3)},
\eqref{(2.4)}
are described in Appendix A. They are eigenstates for the operators
$a_p$ and $a^+_p$:
\beq
\label{(2.5)}
a_p|\phi,\pm\r =\phi_p|\phi,\pm\r\bl{1}(p\in X^\pm),
\eeq
\beq
\label{(2.6)}
\l\phi^*,\pm|a^+_p=\phi^*_p\l\phi^*,\pm|\bl{1}(p\in X^\pm).
\eeq
The scalar product of the coherent states of one type  is given by
the usual formulae \eqref{(A.4)}:
\beq
\label{(2.7)}
\l\phi^*,+|\phi,+\r=\exp\left\{\sl_{p\in X^+}\phi^*_p \phi_p\right\};
\eeq
\beq
\label{(2.8)}
\l\phi^*,-|\phi,-\r=\exp\left\{\sl_{q\in X^-}\phi^*_q \phi_q\right\}.
\eeq
When it cannot cause misunderstandings the sums on the
right hand sides of the equations \eqref{(2.3)}, \eqref{(2.4)},
\eqref{(2.7)}, \eqref{(2.8)} will be denoted $a^+\phi=\sum
a^+_q\phi_q;\phi^+\phi= \sum\phi^+_p\phi_p$ etc.

The scalar products of the coherent states of different type are
given by
\beq
\label{(2.9)}
\l\phi^*,+|\psi,-\r=\exp\left\{\sl_{p\in X^+ ,q\in X^-}\phi^*_p
L_{pq}\psi_q\right\}=\exp\{\phi^*L\psi\},
\eeq
\[\l\psi^*,-|\phi,+\r=\exp\left\{\sl_{p\in X^- ,q\in X^+}\psi^*_p
L_{pq}\phi_q\right\}=\exp\{\psi^*L\phi\},\]
where the matrix element of the $M\times M$ matrix $L$ are
\beq
\label{(2.10)}
L_{pq}=\frac 2M\frac{1}{1-\exp\{-i(p-q)\}}=\frac iM\left(\cot\frac
{q-p}{2}-i\right).
\eeq
It becomes evident if one takes into account that
the Fock vacuum is the same for both types of states
and rewrites  the scalar products in the "coordinate
representation" using the formulae \eqref{(1.21)} and \eqref{(1.22)}.
For example,
\[\l\phi^*,+|\psi,-\r=\l 0|\exp\left\{\sl_{m=1}^M\phi^*_m a_m\right\}
\exp\left\{\sl_{m=1}^M a^+_m\psi_m\right\}|0\r=\]
\beq
\label{(2.11)}
=\exp\left\{\sl_{m=1}^M \phi_m^*\psi_m\right\},
\eeq
where
\[\phi^*_m=\frac{\exp \{i\pi /4\}}{\sqrt{M}}\sl_{p\in X^+} \phi^*_p
\exp\left\{ -i(m-1)p\right\},\]
\beq
\label{(2.12)}
\psi_m=\frac{\exp \{-i\pi /4\}}{\sqrt{M}}\sl_{q\in X^-} \psi_q
\exp\left\{ i(m-1)q\right\}.
\eeq

Consider now the matrix elements of the operator
$\exp \{\a {\bf Q}(m)\}$ where (see \eqref{(1.9)}) ${\bf Q}(m)$ is the
number of particles operator at the first $m$ sites
of the lattice:
\beq
\label{(2.13)}
{\bf Q}=\sl_{l=1}^m a^+_l a_l=\sl_{p_1,p_2}a^+_{p_1}Q_{p_1,p_2}(m)a_{p_2}
\equiv a^+Q(m)a,
\eeq
(the quasimomenta $p_1$ and $p_2$ here correspond to the
states of the same type: $p_1, p_2\in X^+$ or $p_1, p_2\in X^-$).
Using the properties of the matrix $Q(m)$ (evident in the
coordinate representation),
\beq
\label{(2.14)}
Q^2(m)=Q(m); \bl{1}\exp\{\a Q(m)\}=I+(e^\a -1)Q(m),
\eeq
one has for the matrix elements between two states of the
same type
\[\l\phi^*,\pm|\exp\{\a{\bf Q}(m)\}|\phi,\pm\r=\exp\left\{
\phi^*[I+(e^\a -1)Q(m)]\phi\right\}=\]
\beq
\label{(2.15)}
=\exp\left\{\sl_{p\in X^\pm ,q\in X^\pm}\phi^*_p
[\d_{pq}+(e^\a -1)Q_{pq}(m)]\phi_q\right\},
\eeq
the matrix elements of the $M\times M$ matrix $Q(m)$
being given as
\beq
\label{(2.16)}
Q_{pq}(m)=\exp\{ -i(m-1)p/2\}Q_{pq}^{(0)}(m)\exp\{ i(m-1)q/2\},
\eeq
\beq
\label{(2.17)}
Q_{pq}^{(0)}(m)=\frac 1M \frac{\sin \frac{m(p-q)}{2}}{\sin
\frac{p-q}{2}},
\eeq

(for the diagonal matrix elements one should use the l'H\^opital rule,
$Q_{pp}(m)=Q_{pp}^{(0)}(m)=m/M$).

One has for the states of different types, analogously to
\eqref{(2.9)}, \[\l\phi^*,\pm|\exp\{\a{\bf
Q}(m)\}|\phi,\mp\r=\exp\left\{ \phi^*[L+(e^\a -1)Q(m)]\phi\right\}=\]
\beq
\label{(2.18)}
=\exp\left\{\sl_{p\in X^\pm ,q\in X^\mp}\phi^*_p
[L_{pq}+(e^\a -1)Q_{pq}(m)]\phi_q\right\},
\eeq
where the matrix elements $L_{pq}$ and $Q_{pq}(m)$ are given by the
formulae \eqref{(2.10)}, \eqref{(2.16)} (one should note however that
now if $p\in X^+$ then $q\in X^-$ and vice versa). In particular, one
has for the operator $\exp \{i\pi {\bf Q}(m)\}$ entering the
Jordan-Wigner transformation \[\l\phi^*,\pm|\exp\{i\pi{\bf
Q}(m)\}|\phi,\mp\r=\exp\left\{ \phi^*L(m)\phi\right\}=\] \beq
\label{(2.19)}
=\exp\left\{\sl_{p\in X^\pm ,q\in X^\mp}\phi^*_p
L_{pq}(m)\phi_q\right\},
\eeq
where
\beq
\label{(2.20)}
L_{pq}(m)=\exp\{-imp\}L_{pq}\exp\{imq\}.
\eeq

Turn now to the matrix elements ("form factors") of the local
spins. Since
\beq
\label{(2.21)}
{\bf P}^{\pm}\s^\a_m=\s^\a_m{\bf P}^{\mp} \bl{1}(\a=x,y,\bl{1} {\rm or
}\bl{1} \a=\pm),
\eeq
we need only the matrix elements of the local
operators $\s_m^\pm$ between the  states of different type. A direct
calculation using \eqref{(1.7)}, \eqref{(1.22)} and \eqref{(2.18)}
gives :  \[\l\phi^*,\pm|\s^+_m|\psi,\mp\r=\psi_m(\mp)\exp\left\{
\phi^*L(m-1)\psi\right\}=\]
\beq
\label{(2.22)}
=\psi_m(\mp)\exp\left\{\sl_{p\in X^\pm ,q\in X^\mp}\phi^*_p
L_{pq}(m-1)\psi_q\right\},
\eeq
\[\l\phi^*,\pm|\s^-_m|\psi,\mp\r=\phi^*_m(\pm)\exp\left\{
\phi^*L(m-1)\psi\right\},\]
where we use natural notations
\[\psi_m(\pm)=\frac{\exp \{-i\pi /4\}}{\sqrt{M}}\sl_{q\in X^\pm} \psi_q
\exp\left\{ i(m-1)q\right\},\]
\beq
\label{(2.23)}
\phi^*_m(\pm)=\frac{\exp \{i\pi /4\}}{\sqrt{M}}\sl_{p\in X^\pm} \phi^*_p
\exp\left\{ -i(m-1)p\right\}.
\eeq
One should note that the matrices $L$ \eqref{(2.10)}, $L(m)$ \eqref{(2.20)} and $Q(m)$
 \eqref{(2.16)} are related by the following simple formula which can be proved
by direct calculation:
\beq
\label{(2.24)}
\sl_q L_{p_1q}(m)L_{qp_2}=\d_{p_1p_2}-2Q_{p_1p_2}(m),
\eeq
\[(p_1,p_2\in X^+,\bl{1} q\in X^-\bl{1}{\rm or}\bl{1}
(p_1,p_2\in X^-,\bl{1} q\in X^+).\]

Using these formulae one can reproduce the time-dependent temperature
correlation functions for the isotropic XX0 chain obtained in
\cite{cikt1,cikt2,cit}. The
 corresponding calculation is given it in Appendix B. In the next
sections we consider the anisotropic XY chain.

\section{Coherent states for the anisotropic XY chain}
\setcounter{equation}{0}
In this section the formulae for the matrix elements
of operators between the coherent states of the anisotropic
XY chain are given. We consider now two sets of canonical
fermion operators, $a_p,a_p^+$ and $A_p,A_p^+$, related
by the Bogolyubov transformation \eqref{(1.25)}. For each of these sets
we introduce the coherent states ("old" and "new" ones) and
calculate the scalar products of old and new states.

 Consider first the relations between the two vacuum states $|0\r$ and
$|0\r\r$ for the old and new sets
\beq
\label{(3.1)}
a_p|0\r=0, \bl{1}\l a^+_p=0, \bl{1} \forall p;\bl{1}\l0|0\r=1,
\eeq
\beq
\label{(3.2)}
A_p|0\r\r=0, \bl{1}\l\l A^+_p=0, \bl{1} \bl{1}\l\l0|0\r\r=1.
\eeq
 These states are related as follows:
\beq
\label{(3.3)}
|0\r\r=N^{-1/2}\Omega^+|0\r;\bl{1}\l\l0|=N^{-1/2}\l0|\Omega,
\eeq
where the operators $\Omega^+$ and $\Omega$ are
\beq
\label{(3.4)}
\ba{c}
\Omega^+=\exp\left\{\frac 12\sl_p\tau (p)a^+_p a^+_{-p}\right\};\\
\Omega=\exp\left\{\frac 12\sl_p\tau (p)a_{-p} a_{p}\right\},
\ea
\eeq
and
\beq
\label{(3.5)}
\tau (p)\equiv\tan\frac{\theta (p)}{2}.
\eeq
The normalization coefficient $N$ can be represented
as a determinant of
a diagonal $M\times M$ matrix,
\beq
\label{(3.6)}
N=\l 0|\Omega^+\Omega|0\r=\det (I+T),
\eeq
where $I$ is the identity matrix and
\beq
\label{(3.7)}
T=\mbox{diag}(i\tau (p)).
\eeq

Properties  \eqref{(3.2)} for the states \eqref{(3.3)} can be easily checked
using the commutation relations
\[[a_p,\Omega^+]=\tau(p)\Omega^+a^+_{-p}\bl{1} {\rm and}\bl{1}
[\Omega,a_p^+]=\tau(p)a_{-p}\Omega.\]
To calculate the normalization coefficient one makes use of the
equation \eqref{(A.6)} and representing
\[N=\int d\xi
d\xi^*\exp\left\{ -\xi^*\xi +\frac 12 \tau(p)(\xi^*_p\xi^*_{-p}+
\xi_{-p}\xi_p)\right\}.\]
Let us make the change of variables
\beq
\label{(3.8)}
\omega_p=\frac{1}{\sqrt{2}}(\xi_p+i\xi^*_{-p});\bl{1}
\omega^*_p=\frac{1}{\sqrt{2}}(\xi^*_p+i\xi_{-p}).
\eeq
The Jacobian of this transformation is equal to 1, $d\xi d\xi^*=d\omega
d\omega^*$.  One can easily check that \beq \label{(3.9)}
\xi^*\xi=\omega^*\omega;\bl{1}\frac 12\sl_p\tau (p)(\xi^*_p\xi^*_{-p}+
\\xi_{-p}\xi_p)=-\omega^*T\omega,
\eeq
where the matrix $T$ is defined in \eqref{(3.7)}. Hence, $N=\int d\omega d\omega^*
\exp\{-\omega^*(I+T)\omega\}=\det (I+T)$, which proves the relation \eqref{(3.6)}.

 Introduce now the coherent states for the new set  of operators
(compare with \eqref{(A.2)}):
\[|X\r\r=\exp\left\{\sl_pA^+_pX_p\right\}|0\r\r;\]
\beq
\label{(3.10)}
\l\l X^*|=\l\l0|\exp\left\{\sl_p X^*_p A_p\right\}.
\eeq
The Grassman algebra elements $X_p, X_p^*$ have the same properties
as the old parameters $\xi_p,\xi_p^*$. The formulae
\eqref{(A.2)}-\eqref{(A.10)}
are valid, of course, also for the new states. The direct calculation
gives the following representations for the scalar products of old and
new states
\[\l\xi^*|X\r\r=N^{-1/2}\exp\left\{\frac
12\sl_p\tau (p)(\xi^*_p \xi^*_{-p}-
X_{-p}X_p)+\sl_p\a^{-1}(p)\xi^*_pX_p\right\},\]
\beq \label{(3.11)} \l\l
X^*|\xi\r=N^{-1/2}\exp\left\{\frac 12\sl_p\tau (p)( \xi_{-p}\xi_p-
X^*_pX^*_{-p})+\sl_p\a^{-1}(p)X^*_p\xi_p\right\}.\eeq
Considering the XY
 model it is necessary to introduce different sets of operators
$a_p,a_p^+$ and $A_p,A_p^+$ corresponding to the sets of momenta  $X_+$
(for ${\bf H^+}$) and $X_-$ (for ${\bf H^-}$), see
\eqref{(1.19)},\eqref{(1.20)}. Thus the new vacuums $|\Omega\r\r$ will
be also different for ${\bf H^+}$ and ${\bf H^-}$ (unlike the
old vacuum $|0\r$). We will not usually mark
in our notations this difference but one should have it in mind.

 Using the equation \eqref{(3.11)}, one can calculate the matrix
elements of the operators $\exp\{-\b {\bf H^\pm}\}$ diagonal in the new
representations (different for ${\bf H^+}$ and
${\bf H^-}$!) between old coherent states
(also different for ${\bf H^+}$ and ${\bf H^-}$);
\beq
\label{(3.12)}
\l\xi^*|e^{-\b {\bf H^\pm}}|\xi\r=N^{-1}e^{-\b E_0^\pm}\det (I-J_\b T)
e^{\omega^*D\omega},
\eeq
where the variables $\omega_p, \omega_p^*$ ($p\in X^+$ for ${\bf H^+}$
and $p\in X^-$ for ${\bf H^-}$) are defined by the equation
\eqref{(3.8)}, and diagonal $M\times M$ matrices $J_\b$ and $D$ are
\beq \label{(3.13)} J_\b=\mbox{diag}(j_\b (p));\bl{1} j_\b
(p)=\exp\{-\b E(p)\}, \eeq
\beq \label{(3.14)} D=\mbox{diag}
(d(p));\bl{1} d(p)= \frac{j_\b (p)-i\tau (p)}{1-i\tau(p)j_\b (p)}, \eeq
(the matrix $T$ is defined in \eqref{(3.7)}).

 To prove it we use the completeness of the new states \eqref{(A.6)},
\[\l\xi^*|e^{-\b {\bf H^\pm}}|\xi\r=\int dX dX^* dY dY^*\times\]
\[\times\l\xi^*|X\r\r\l\l Y^*|e^{-\b {\bf H^\pm}}|Y\r\r\l\l X^*|\xi\r
\exp\{-Y^*X-X^*Y\},\]
and note that the operators ${\bf H^\pm}$ are diagonal in the corresponding
new representations \eqref{(1.31)}:
\beq
\label{(3.15)}
\l\l Y^*|e^{-\b {\bf H^\pm}}|Y\r\r=e^{-\b E_0^\pm}\exp\left\{\sl_{p\in X^\pm}
e^{-\b E(p)}Y^*_pY_p\right\}.
\eeq
After the integration over $Y, Y^*$ one gets
\[\l\xi^*|e^{-\b {\bf H^\pm}}|\xi\r=\int dX dX^*\l\xi^*|X\r\r\l\l X^*|\xi\r\times\]
\[\times e^{-\b E_0^\pm}\exp\left\{\sl_{p\in X^\pm}e^{-\b E(p) X^*_pX_p}
\right\}.\]
 The integral over $X, X^*$ can be calculated by means of the change of
variables as in \eqref{(3.8)}. Finally one gets the formula
\eqref{(3.12)}.

 In the following sections the results obtained here will be used to
calculate the equal-time correlators.

\section{ The simplest correlator for the XY chain}
\setcounter{equation}{0}
 In this section the partition function $Z$ and the
generating functional of the third components of local spins
are calculated for the anisotropic chain. We begin
by considering the partition function. The initial representation
is the same as in th isotropic case \eqref{(B.1)}
\beq
\label{(4.1)}
Z=\frac 12 (Z_F^++Z_F^-+Z_B^+-Z_B^-),
\eeq
where
\beq
\label{(4.2)}
\ba{c}
Z_F^\pm=\mbox{Tr}\exp\{-\b{\bf H^\pm}\};\\
Z_B^\pm=\mbox{Tr}(\exp\{-\b{\bf H^\pm}\}(-1)^{\bf N}).
\ea
\eeq
One gets the following representations for the contributions
(see for example \cite{ki}; the $M \times M $ matrix $J_\b$ is
defined in \eqref{(3.13)}):
\beq
\label{(4.3)}
Z_F^\pm=e^{-\b E_0^\pm}\det (I+J_\b )=\pl_{p\in X^\pm}\left(
2\cosh\frac{\b E(p)}{2}\right),
\eeq
\beq
\label{(4.4)}
Z_B^\pm=e^{-\b E_0^\pm}\det (I-J_\b )=\pl_{p\in X^\pm}\left(
2\sinh\frac{\b E(p)}{2}\right).
\eeq
 To obtain, e.g., the fermionic contributions one
should use the representation \eqref{(A.5)} for the trace of
operators:
\[Z_F^\pm=\int dY dY^* \l\l Y^*|e^{-\b {\bf H^\pm}}|Y\r\r\exp\{Y^*Y\},\]
and the representation \eqref{(3.15)} for the matrix element involved;
after that one should  calculate the Gaussian integral  leading to
the equality \eqref{(4.3)}. The "bosonic" contrubutions can be
calculated analogously.

 It is worth mentioning that it is sometimes convenient  to represent
the answer differently using the old coherent states, for example,
\[Z_F^\pm=\int d\xi d\xi^*\l\xi^*|e^{-\b {\bf
H^\pm}}|\xi\r\exp\{\xi^*\xi\}. \]
By means of equation \eqref{(3.12)} for the
corresponding matrix element one gets
\beq \label{(4.5)}
Z_F^\pm=N^{-1}e^{-\b E_0^\pm}\det (I-J_\b T)\det (I+D),
\eeq
\beq
\label{(4.6)}
Z_B^\pm=N^{-1}e^{-\b E_0^\pm}\det (I-J_\b T)\det (I-D).
\eeq

 Turn now to the simplest equal-time temperature correlator
\beq
\label{(4.7)}
G(M)=\frac 1Z\mbox{Tr}\left(e^{\a{\bf Q}(m)}e^{-\b{\bf H}}\right).
\eeq
As for the isotropic chain (\eqref{(B.8)},\eqref{(B.9)}) it can be represented as
a sum of four contributions
\beq
\label{(4.8)}
G(m)=\frac 1{2Z} (Z_F^+G_F^++Z_F^-G_F^-+Z_B^+G_B^+-Z_B^-G_B^-),
\eeq
where
\[Z_F^\pm G_F^\pm=\mbox{Tr}\left(e^{\a{\bf Q}(m)}e^{-\b{\bf
H^\pm}}\right),\]
\beq \label{(4.9)} Z_B^\pm
G_B^\pm=\mbox{Tr}\left(e^{\a{\bf Q}(m)}e^{-\b{\bf H^\pm}}(-1)^{\bf N}
\right).
\eeq

 One has for the contributions $G_{F,B}^\pm$ the following representations
as determinants of $M\times M$ matrices
\beq
\label{(4.10)}
G_F^\pm=\det\left(I+(e^\a-1)Q^{(0)}(m)\Omega_F\right),
\eeq
\beq
\label{(4.11)}
G_B^\pm=\det\left(I-(e^\a-1)Q^{(0)}(m)\Omega_B\right).
\eeq
Here the matrix elements of $Q^{(0)}(m)$ are given by \eqref{(2.16)}
with $p,q\in X^+$ for $G^+_{F,B}$ and $p,q\in X^-$ for $G^-_{F,B}$.
 Diagonal $M\times M$ matrices $\Omega_F$ and  $\Omega_B$ are given
by the following formulae ($p\in X^\pm$ for $G^\pm_{F,B}$)
\[
\Omega_F=D(I+D)^{-1}=\mbox{diag}(\omega_F(p));\]
\beq
\label{(4.12)}
\omega_F(p)=\frac 12 \left( 1-e^{i\theta (p)}\tanh\frac{\b E(p)}{2}\right),
\eeq
\[\Omega_B=D(I-D)^{-1}=\mbox{diag}(\omega_B(p));\]
\beq
\label{(4.13)}
\omega_B(p)=\frac 12 \left( e^{i\theta (p)}\coth\frac{\b E(p)}{2}-1\right),
\eeq
(the matrix $D$ is defined in \eqref{(3.14)}).

 Explain now how to calculate the contribution $G_F^+$:
\beq
\label{(4.14)}
Z_F^+G_F^+=\int d\xi d\xi^*d\eta d\eta^*\l\eta^*|e^{\a{\bf Q}(m)}|\eta\r
\l\xi^*|e^{-\b{\bf H^+}}|\xi\r e^{\eta^*\xi-\xi^*\eta}.
\eeq
The matrix elements on the right hand side are given by \eqref{(2.15)}
and \eqref{(3.12)}. We change the variables $\eta^*\rightarrow -\eta^*$;
since $M$ is even, the integration measure is invariant. After that
we change the variables as in \eqref{(3.8)}
\beq
\label{(4.15)}
\ba{c}
\omega_p=\frac{1}{\sqrt{2}}(\xi_p+i\xi^*_{-p});\bl{1}
\omega^*_p=\frac{1}{\sqrt{2}}(\xi^*_p+i\xi_{-p}),\\
\rho_p=\frac{1}{\sqrt{2}}(\eta_p+i\eta^*_{-p});\bl{1}
\rho^*_p=\frac{1}{\sqrt{2}}(\eta^*_p+i\eta_{-p}).
\ea
\eeq
 As a result of this change of variables the measure remains
invariant and
\[\eta^*\xi+\xi^*\eta=\rho^*\omega+\omega^*\rho,\]
\beq
\label{(4.16)}
\eta^*(I+(e^\a-1)Q(m))\eta=\rho^*(I+(e^\a-1)Q(m))\rho.
\eeq
 The integration over $\omega,\omega^*$ of the factors depending
on these variables in \eqref{(4.14)} gives $\det D\exp\{-\rho^*D^{-1}\rho\}$.
One can calculate the remaining Gauss integral on $\rho, \rho^*$
taking into account the representation \eqref{(4.5)} for the partition
functions and equation \eqref{(4.12)}:
\beq
\label{(4.17)}
Z_F^+G_F^+=Z_F^+\det(I+(e^\a-1)Q(m)\Omega_F).
\eeq
One should note that the matrix $Q(m)$ \eqref{(2.16)} differs only by an
evident similarity transformation with a diagonal matrix from the
matrix $Q^{(0)}(m)$. Thus the representation \eqref{(4.10)} for $G_F^+$ is
proved. The derivation of the contribution $G_F^-$ is almost the same,
one should only use the momenta from the set $X^-$. To calculate the
bosonic contributions one should use the
 property of the operator $(-1)^{\bf N}$,
\beq
\label{(4.18)}
(-1)^{\bf N}|\xi\r=|-\xi\r,
\eeq
in the representation similar to \eqref{(4.14)} which change evidently
the calculations leading to the result \eqref{(4.11)}.

 We should make a remark about the equations \eqref{(4.10)},
\eqref{(4.11)}.  Since the sets $X^+$ and $X^-$ are symmetric under the
change of momenta $p_i\rightarrow -p_i$ (there is an exception, it is
the momenta $0$ and $\pi$ from the set $X^-$; but the result is valid
also in this case) one can rewrite the answer  as (the sign "+" on the
right hand side corresponds to $G_F$ and the sign "-" corresponds to
$G_B$)
\[G^\pm_{F,B}=\det(I\pm(e^\a-1)Q(m)\Omega_{F,B})=\] \[=\det(I\pm
(e^\a-1)\Omega^{1/2}_{F,B}Q(m)\Omega^{1/2}_{F,B})=\]
\beq
\label{(4.19)}
=\det(I\pm
(e^\a-1)\bar{\Omega}^{1/2}_{F,B}Q(m)\bar{\Omega}^{1/2}_{F,B})= \eeq
\[=\det(I\pm(e^\a-1)Q(m)\bar{\Omega}_{F,B}).\]
The bar means here the complex conjugation and $\omega_{F,B}(-p)=
\bar{\omega}_{F,B}(p)$

To conclude we discuss some limiting cases.

In the zero temperature limit ($\b\rightarrow\infty$) the "odd" contributions
to the correlator are cancelled and one gets the representation for $G(m)$
\beq
\label{(4.20)}
G(m)=\det(I+(e^\a-1)Q^{(0)}(m)\Omega_0)\bl{1} (T=0),
\eeq
where
\beq
\label{(4.21)}
\Omega_0=\mbox{diag}(\omega_0(p)),\bl{1} \omega_0(p)=\frac 12
\left(1-e^{i\theta(p)}\right),
\eeq
coinciding with the result obtained in \cite{iks}. On the other hand
for the isotropic case ($\g =0$) taking into account that the angle
$\theta (p)$ in the Bogolyubov transformation is
\beq
\label{(4.22)}
\ba{c}
\theta(p)=-\pi\mbox{sign}p,\bl{1}|p|<k_F;\\
\theta(p)=0,\bl{1}|p|>k_F\bl{1} (\g=0),
\ea
\eeq
one gets for the weights
\[\omega_F(p)=\frac 12 \left( 1-\tanh\frac{\b
\varepsilon(p)}{2}\right),\]
\beq
\label{(4.23)}
\omega_B(p)=\frac 12 \left( \coth\frac{\b
\varepsilon(p)}{2}-1\right) \bl{1} (\g=0).
\eeq
Here $k_F=\arccos
h$ is the Fermi momentum and $\varepsilon (p)$ is the dispersion of the
XX0 chain (see \eqref{(1.24)}). Thus  one reproduces  the answers for
the isotropic case \cite{cikt1,cit}.

\section{Equal-time correlators of the local spins}
\setcounter{equation}{0}
Here we consider the equal-time correlation functions of
the local spin operators ($\b\equiv 1/T$):
\beq
\label{(5.1)}
G^{(ab)}(m)=\l\s^a_{m+1}\s^b_1\r_T=\frac 1Z \mbox{Tr}(\s^a_{m+1}\s^b_1
e^{-\b{\bf H}}),\bl{1} a,b=+,-.
\eeq
These correlators  on a finite lattice of length $M$ can be represented
(as in \eqref{(B.8)}) as a sum of four contributions
\beq
\label{(5.2)}
G^{(ab)}=\frac 1{2Z} \left(Z_F^+G_F^{(ab),+}+Z_F^-G_F^{(ab),-}
+Z_B^+G_B^{(ab),+}-Z_B^-G_B^{(ab),-}\right),
\eeq
where
\[Z_F^\pm G_F^{(ab),\pm}=\mbox{Tr}\left(\s^a_{m+1}\s^b_1e^{-\b{\bf
H^\pm}}\right),\]
\beq
\label{(5.3)}
Z_B^\pm
G_B^{(ab),\pm}=\mbox{Tr}\left(\s^a_{m+1}\s^b_1e^{-\b{\bf
H^\pm}}(-1)^{\bf N} \right),
\eeq
(partition functions
$Z_{F,B}^\pm$ are given by \eqref{(4.3)}, \eqref{(4.4)}). For the
contributions we obtain the following representations as determinants
of $M\times M$ matrices:
\beq
\label{(5.4)}
G_F^{(-+),\pm}=G_F^{(+-),\pm}=\left.\frac{\partial}{\partial\a}\det (I+U\Omega_F
+\a C\Omega_F)\right|_{\a=0}\bl{1}(m>0),
\eeq
\[
G_F^{(-+),\pm}=\frac 1M \mbox{tr}\Omega_F=\frac 1M\sl_p\omega_F(p)\bl{1}
(m=0);\]
\beq
\label{(5.5)}
G_F^{(+-),\pm}=1-\frac 1M \mbox{tr}\Omega_F\bl{1}
(m=0),
\eeq
\beq
\label{(5.6)}
G_F^{(++),\pm}=G_F^{(--),\pm}=\left.\frac{\partial}{\partial\a}\det (I+U\Omega_F
-i\a S\Omega_F)\right|_{\a=0}\bl{1}(m\ge 0),
\eeq
\beq
\label{(5.7)}
G_B^{(-+),\pm}=G_B^{(+-),\pm}=\left.\frac{\partial}{\partial\a}\det (I-U\Omega_F
-\a C\Omega_F)\right|_{\a=0}\bl{1}(m>0),
\eeq
\beq
\label{(5.8)}
G_B^{(-+),\pm}=-\frac 1M \mbox{tr}\Omega_B;\bl{1}
G_F^{(+-),\pm}=1+\frac 1M \mbox{tr}\Omega_B\bl{1} (m=0),
\eeq
\beq
\label{(5.9)}
G_B^{(++),\pm}=G_B^{(--),\pm}=\left.\frac{\partial}{\partial\a}\det (I-U\Omega_F
-i\a S\Omega_F)\right|_{\a=0}\bl{1}(m\ge 0).
\eeq
The diagonal $M\times M$ weight matrices $\Omega_F$ and $\Omega_B$ are
defined in \eqref{(4.12)}, \eqref{(4.13)}. Matrix elements of $M\times
M$ matrices $U,C,S$ are given by the following formulae \beq
\label{(5.10)}
U_{p_1p_2}=-\exp\left\{\frac i2(p_1-p_2)\right\}Q^{(0)}_{p_1,p_2}(m),
\eeq
(the definition of the matrix $Q^{(0)}(m)$ is in \eqref{(2.17)}),
\beq
\label{(5.11)}
C_{p_1p_2}=\frac 1M\cos\frac m2(p_1+p_2);
\eeq
\beq
\label{(5.12)}
S_{p_1p_2}=\frac 1M\sin\frac m2(p_1+p_2).
\eeq
It is necessary to emphasize that the momenta numerating the matrix elements
$p_1,p_2\in X^+$ for the contributions $G_{F,B}^{(ab),+}$ and
$p_1,p_2\in X^-$ for the contributions $G_{F,B}^{(ab),-}$; the same
thing is true for the weight matrices.

Calculating the functions $G_{F,B}^{(ab),\pm}$ is reduced to calculating
Gaussian integrals in the Grassmanian variables. For example, using
 \eqref{(A.5)} and \eqref{(A.6)} we represent
\beq
\label{(5.13)}
Z_F^+G_F^{(ab)+}=\int d\xi d\xi^*d\eta d\eta^*\l\eta^*|
\sigma^a_{m+1}\sigma_1^b|\eta\r
\l\xi^*|e^{-\b{\bf H^+}}|\xi\r e^{\eta^*\xi-\xi^*\eta}.
\eeq
Consider the calculation of the matrix element
\[F_{ab}=\l\eta^*|\sigma^a_{m+1}\sigma_1^b|\eta\r=\]
\[=\int d\zeta d\zeta^*\l\eta^*|\sigma^a_{m+1}|\zeta\r\l\zeta^*|
\sigma_1^b|\eta\r e^{-\zeta^*\zeta}.\]
Let us use the formulae \eqref{(2.22)} for the matrix elements of spin
operators and make the change of variables \beq \label{(5.14)} \ba{c}
\tilde{\eta}_p^*=e^{-\frac{im}{2}p}\eta^*_p;\bl{1}
\tilde{\eta}_p=e^{\frac{im}{2}p}\eta_p\bl{1}(p\in X^+),\\
\tilde{\zeta}_q^*=e^{-\frac{im}{2}q}\zeta^*_q;\bl{1}
\tilde{\zeta}_q=e^{\frac{im}{2}q}\zeta_q\bl{1}(q\in X^-).
\ea
\eeq
One gets the representation for the matrix element using the new variables
(we omit tildes over the new variables \eqref{(5.14)}):
\beq
\label{(5.15)}
F_{ab}=\frac{\partial}{\partial\a}\int \left.d\zeta d\zeta^*\exp\{\omega+\a
f_{ab}\}\right|_{\a=0},
\eeq
where
\beq
\label{(5.16)}
\omega=\eta^*PL\bar{P}\zeta+\zeta^*\bar{P}LP\eta-\zeta^*\zeta,
\eeq
\beq
\label{(5.17)}
f_{-+}=\eta^*R_+\eta;\bl{1}(R_+)_{p_1p_2}=
\frac 1M e^{-\frac{im}{2}(p_1+p_2)},
\eeq
\beq
\label{(5.18)}
f_{+-}=\zeta^*\bar{R}_+\zeta;\bl{1}(\bar{R}_+)_{q_1q_2}=
\frac 1M e^{\frac{im}{2}(q_1+q_2)},
\eeq
\beq
\label{(5.19)}
f_{++}=i\eta R_-\zeta;\bl{1}(R_-)_{pq}=
\frac 1M e^{-\frac{im}{2}(p-q)},
\eeq
\beq
\label{(5.20)}
f_{--}=i\zeta^* \bar{R}_-\eta^*;\bl{1}(\bar{R}_-)_{qp}=
\frac 1M e^{\frac{im}{2}(q-p)}=(R_-)_{pq},
\eeq
(the bar means the complex conjugation). The diagonal
matrix $P$ has the form
\beq
\label{(5.21)}
P=\mbox{diag}\left(e^{-\frac{im}{2}p}\right).
\eeq

Use now the following identities
\[\sl_q L_{pq}e^{imq}=(2\d_{m,0}-1)e^{ipm};\]
\beq
\label{(5.22)}
\sl_q e^{imq} L_{qp}=e^{ipm} \bl{1} (p\in X^+,q\in X^-,
\bl{1}m=0,1,\dots,M-1),
\eeq
\beq
\label{(5.23)}
(PL\bar{P}\bar{R}_+\bar{P}LP)_{p_1p_2}=-(\bar{R}_+)_{p_1p_2}(1-2\d_{m,0})
\eeq
\beq
\label{(5.24)}
(R_-\bar{P}LP)_{p_1p_2}=(R_-)_{p_1p_2}
\eeq
\beq
\label{(5.25)}
(PL\bar{P}\bar{R}_-)_{p_1p_2}=-(1-2\d_{m,0})(\bar{R}_-)_{p_1p_2}
\eeq
It should be emphasized that the matrix
elements of the matrices on the left hand side
of \eqref{(5.23)} and \eqref{(5.24)} are
\[(\bar{R}_+)_{q_1q_2}(q_{1,2}\in X^-)\bl{1}
{\rm and} \bl{1}
(R_-)_{pq}(p\in X^+,q\in X^-),\]
and on the right hand side $(\bar{R}_+)_{p_1p_2}$,
$(R_-)_{p_1p_2}(p_{1,2}\in X^+)$. As a result,
the matrix elements $F_{ab}$ \eqref{(5.15)} are represented
as follows
\[F_{-+}=\left.\pd{\a}\exp\{\eta^*(I+U)\eta+\a\eta^*R_+\eta\}\right|_{\a=0},\]
\[F_{+-}=\left.(\d_{m,0}+\pd{\a})\exp\{\eta^*(I+U)\eta+
\a\eta^*\tilde{\bar{R}}_+\eta\}\right|_{\a=0},\]
\beq
\label{(5.26)}
F_{++}=\left.\pd{\a}\exp\{\eta^*(I+U)\eta+\a\eta^*S_-\eta\}\right|_{\a=0},
\eeq
\[F_{--}=\left.\pd{\a}\exp\{\eta^*(I+U)\eta-\a\eta^*S_-\eta\}\right|_{\a=0}.\]
Here the matrix elements of the $M\times M$ matrix $U$ are
defined as $U_{p_1p_2}=(PL\bar{P}^2LP)_{p_1p_2}-\delta_{p_1p_2}$
and this representation leads to \eqref{(5.10)} if
one takes into account \eqref{(2.24)}; the  matrices
$R_+$, $\bar{R}_+$ are given by \eqref{(5.17)}, \eqref{(5.18)} and \eqref{(5.12)};
the matrix $\tilde{\bar{R}_+}$ is defined by the relation
\beq
\label{(5.27)}
(\tilde{\bar{R}}_+)_{p_1p_2}=(\bar{R}_+)_{p_1p_2}(1-2\d_{m,0}),
\eeq
\beq
\label{(5.28)}
(S_-)_{p_1p_2}=\frac 1M\sin\frac m2 (p_1-p_2).
\eeq
 Now we put the expressions \eqref{(5.12)} into \eqref{(5.3)} using also \eqref{(3.12)}.
Then it is not difficult to get for the contribution $G_F^{(-+),+}$
\[ Z_F^+G_F^{(-+),+}=N^{-1}e^{-\b E_0^+}\det(I-J_\b T)\pd{\a}\int d\xi d\xi^*
d\eta d\eta^*\times\]
\beq
\label{(5.29)}
\times\exp\{-\eta^*(I+U+\a R_+)\eta+\omega^*D\omega-
\eta^*\xi-\xi^*\eta\},
\eeq
(besides \eqref{(3.8)}, we changed the variables as in \eqref{(5.14)}
and also we changed $\eta^*\rightarrow -\eta^*$). Finally we change the
variables as in \eqref{(4.15)}, $(\eta ,\eta^*)\rightarrow(\rho
,\rho^*)$; $(\xi, \xi^*) \rightarrow (\omega, \omega^*)$. Using
\eqref{(4.16)} and equalities \[\eta^*(I+U+\a R_+)\eta =\rho^*(I+U+\a
C)\rho -\frac{\a}{2}(\rho^* S_-\rho^*-\rho S_-\rho),\] we integrate
over $\omega, \omega^*$ and then over $\rho ,\rho^*$ in \eqref{(5.29)}.
As a result we get the following representation as a determinant of
a $2M\times 2M$ matrix:
\[ Z_F^+ G_F^{(-+),+}=N^{-1} e^{-\b E_0^+}\det (I-J_\b T)\det D\times\]
\[\times\frac{\partial}{\partial\a}\det^{1/2}\left(\ba{ccc}
\a S_- & \vdots & -B \\ \cdots & \cdots & \cdots \\ B^T & \vdots & -\a S_-
\ea\right),\]
where we denoted $B=I+D^{-1}+U+\a C$. Using the generalized
Gauss algorithm it is not difficult to check that
\[\det\left(\ba{ccc}
\a S_- & \vdots & -B \\ \cdots & \cdots & \cdots \\ B^T & \vdots & -\a S_-
\ea\right)=\det^2 B+O(\a^2),\]
(it is valid also for $m$=0). It leads to the representation for
$G_F^{(-+),+}$. Analogously one can calculate other contributions.

One should note that the expressions \eqref{(5.4)}-\eqref{(5.9)} for equal-time
temperature correlators of the anisotropic XY chain differ
from the corresponding expressions for the isotropic XX0
chain only by changing the weights. For example,  the corresponding modification
for the "fermionic" weightis as before
\beq
\label{(5.30)}
\frac 12 \left(1-\tanh\frac{\b\varepsilon (p)}{2}\right)\rightarrow
\frac 12 \left(1-e^{i\theta (p)}\tanh\frac{\b E(p)}{2}\right).
\eeq

\section{The thermodynamic limit}
\setcounter{equation}{0}
Now we consider the correlators for the anisotropic XY chain
in the thermodynamic limit (the length of the chain $M\rightarrow\infty$,
the magnetic field $h$ is fixed). In this limit the partition functions
$Z_{F,B}^\pm$ are divergent and using \eqref{(4.3)}, \eqref{(4.4)} one
can estimate
\[Z_B/Z_F <C^M; \bl{1} C=\tanh \frac{\b E_{\rm
max}}{2}<1,\]
where $E_{\rm max}$ is the maximal value of the quasiparticle energy
$E(p)$ \eqref{(1.28)}. So only the "fermionic" contributions survive in
the thermodynamic limit in \eqref{(4.8)} and \eqref{(5.2)}. The
determinants of the $M\times M$ matrices in the expressions
\eqref{(4.10)}, \eqref{(5.4)} for these contributions become in the
thermodynamic limit the Fredholm determinants of the corresponding
integral operators. It is explained explicitly,
for example, in \cite{cikt2}. So we get the following
answers.

The correlator $G(m)$ \eqref{(4.7)} is given by
\beq
\label{(6.1)}
G(m)=\lim\frac 1Z\mbox{Tr}\left(e^{\a{\bf Q}(m)}e^{-\b{\bf H}}\right)
=\det\left(\hat{I}+(e^\a -1)\hat{V}\right).
\eeq
At the right hand side, there is the Fredholm determinant of a linear
integral operator acting on functions $f(p)$ on the interval $-\pi\le
p\le\pi$ \beq \label{(6.2)}
(\hat{V}f)(p)=\int\limits_{-\pi}^{\pi}V(p,q)f(q)dq,
\eeq
$\hat{I}$ means the identity operator on the interval and
the kernel $V(p,q)$ is (see \eqref{(4.10)}, \eqref{(2.17)}, \eqref{(4.12)})
\beq
\label{(6.3)}
V(p,q)=\frac {1}{2\pi}\frac{\sin\frac{m(p-q)}{2}}{\sin\frac{p-q}{2}}.
\omega_F(q),
\eeq
The weight $\omega_F(q)$ is equal to
\beq
\label{(6.4)}
\omega_F(q)=\frac 12\left(1-e^{i\theta (q)}\tanh\frac{\beta E(q)}{2}\right)
\eeq
(the angle $\theta(p)$ and the quasiparticle energy $E(p)$ are
defined in  \eqref{(1.27)}, \eqref{(1.28)}).

In the thermodynamic limit the correlators \eqref{(5.1)}
\beq
\label{(6.5)}
G^{(ab)}(m)=\lim\frac 1Z \mbox{Tr}(\s^a_{m+1}\s^b_1
e^{-\b{\bf H}})
\eeq
can be also represented as Fredholm determinants of linear
integral operators on the interval $[-\pi,\pi ]$. From
 \eqref{(5.4)}-\eqref{(5.6)} one gets for $m>0$:
\beq
\label{(6.6)}
G^{(-+)}(m)=G^{(+-)}(m)=\left.\pd{\a}\det (\hat{I}-\hat{W}+\a\hat{C})
\right|_{\a=0}\bl{1}(m>0),
\eeq
\beq
\label{(6.7)}
G^{(++)}(m)=G^{(--)}(m)=\left.\pd{\a}\det (\hat{I}-\hat{W}-i\a\hat{S})
\right|_{\a=0}\bl{1}(m>0).
\eeq
The kernels of the operators $\hat{W}$, $\hat{C}$ and $\hat{S}$ are
\beq
\label{(6.8)}
W(p,q)=-\frac 1\pi e^\frac{i(p-q)}{2}
\frac{\sin\frac{m(p-q)}{2}}{\sin\frac{p-q}{2}}\omega_F(q),
\eeq
\beq
\label{(6.9)}
C(p,q)=\frac{1}{2\pi}\cos\frac{m(p+q)}{2}\omega_F(q)
\eeq
\beq
\label{(6.10)}
S(p,q)=\frac{1}{2\pi}\sin\frac{m(p+q)}{2}\omega_F(q).
\eeq
 These answers differ only by changing the weight \eqref{(5.30)} from the
answers for the isotropic XX0 chain.

 One should note that the formal answers for the time-dependent
correlation functions were obtained recently in \cite{ika}.
In our next paper we hope to give more clear answers for
the time-dependent correlators.

 We are grateful to the Russian Foundation of Fundamental Research
(grants 95-01-00476 and 96-01-00807) and INTAS (grant INTAS-RFFR 95-414).
One of us (V.S.K.) thanks also TOO "Cyclone" for the support.

\section*{Appendix A}
\renewcommand{\theequation}{A.\arabic{equation}}
\setcounter{equation}{0}
 Consider a set of canonical fermion operators
\[a_q, a_q^+\bl{1} ([a_p,a_q]_+=[a^+_p,a^+_q]_+=0, [a_p, a_q^+]_+=\d_{p,q}).\]
We denote by $M$ the number of operators $a$ (or $a^+$) in the set.
It is supposed that the Fock vacuum exists and has the following properties
\beq
\label{(A.1)}
a_q|0\r=0;\bl{1}\l0| a^+_q=0,\bl{1}\forall q;\bl{1}\l0|0\r=1.
\eeq
 We introduce the coherent states \cite{bie}
\[ |\xi\r\equiv |\xi_q\r=\exp\left\{\sl_qa^+_q\xi_q\right\}|0\r;\]
\beq
\label{(A.2)}
\l\xi^*|\equiv \l\xi_q^*|=\l 0|\exp\left\{\sl_q\xi^*_q a_q\right\},
\eeq
where the summation is taken over the whole set. The parameters
$\xi_q,\xi_q^+$  (Grassman algebra elements) anticommute among
themselves and with all the operators $a_q, a_q^+$. One should
emphasize that the star in $\xi^*$ means only that the corresponding
parameter is connected with a bra-vector; we don't consider
involutions on the Grassman algebra; parameters $\xi$ and $\xi^*$
are entirely independent. The coherent states \eqref{(A.2)} are
the eigenstates for the creation and annihilation
operators, respectively
\beq \label{(A.3)}
a_p|\xi\r=\xi_p|\xi\r;\bl{1}\l\xi^*|a^+_p=\l\xi^*|\xi^*_p.
\eeq
One can easily calculate the scalar product of two coherent states
using the commutation relations between $a$, $a^+$, $\xi$, $\xi^*$:
\beq
\label{(A.4)}
\l\xi^*|\xi\r=\exp\left\{\sl_q\xi_q^*\xi_q\right\}\equiv\exp\{\xi^*\xi\}.
\eeq
The trace of an  operator  ${\bf O}$ can be represented
as an integral over the anticommuting variables
of the matrix elements of the operator between the
coherent states (see \cite{bie}):
\beq
\label{(A.5)}
\mbox{Tr}{\bf O}=\int d\xi d\xi^*\exp\{\xi^*\xi\}\l\xi^*|{\bf O}|\xi\r,
\eeq
and the expansion for the identity operator is given by
\beq
\label{(A.6)}
{\bf 1}=(-1)^M\int d\xi d\xi^*\exp\{-\xi^*\xi\}|\xi\r\l\xi^*|,
\eeq
(we supposed that the number of sites $M$  is  even and the coefficient
$(-1)^M$ in such formulae is usually omitted). We use the following
notation \[d\xi\equiv\pl_q d\xi_q;\bl{1} d\xi^*\equiv\pl_q d\xi^*_q.\]

If an operator ${\bf L}$ has the following form
\[{\bf L}=\sl_{p,q}a^+_p L_{pq} a_q\equiv a^+ La,\]
where $L$ is a $M\times M$ matrix (with matrix elements $L_{pq}$)
which can be diagonalized by an unitary matrix $U$,
\[L=U^+DU;\bl{1}U^+U=UU^+=I;\bl{1} D=\mbox{diag}(D_q),\]
then the matrix elements of the operator $\exp{\bf L}$ are given by
\beq
\label{(A.7)}
\l\xi^*|\exp\{{\bf L}\}|\xi\r=\exp\left\{\sl_{p,q}\xi^*_p(\exp \{L\})_{pq}
\xi_q\right\}\equiv\exp\{\xi^*\exp L\xi\},
\eeq
and
\beq
\label{(A.8)}
\mbox{Tr}\exp\{{\bf L}\}=\mbox{Tr}\exp\{a^+La\}=\det [I+\exp\{L\} ],
\eeq
(there is the trace of the operator on the left hand side
of the last formula and the  determinant of the $M\times M$ matrix
on the right hand side).

The last equality follows from a well-known formula for the Gaussian
integral over the anticommuting variables
\beq
\label{(A.9)}
\int d\xi d\xi^*\exp\{\xi^* K\xi \eta^*\xi+\xi^*\eta\}=
\exp\{-\eta^*K^{-1}\eta\}\det K.
\eeq
We need also to use a formula valid for an antisymmetric matrix
$A$,
\beq
\label{(A.10)}
\int d\xi\exp\left\{\frac 12 \xi A\xi +\eta\xi\right\}=
\exp\left\{\frac 12 \eta A^{-1}\eta\right\}\sqrt{\det A},\bl{1} A=-A^T.
\eeq

\section*{Appendix B}
\renewcommand{\theequation}{B.\arabic{equation}}
\setcounter{equation}{0}
Here we give the derivation of the formulae for the time-dependent
correlators for the isotropic XX0 chain with finite number of sites
$M$ ($M$ is supposed to be even). We hope that readers will
appreciate the simplicity  of the derivation of the results
using the integration over the Grassmanian variables
(compared
with paper \cite{cikt2,cit}). Below ${\bf H}$ denotes the Hamiltonian
of the XX0 model We begin  with the calculation of the partition
function $Z$.  Following \cite{McCBA,ki} we represent it in the form
\beq
\label{(B.1)}
Z=\mbox{Tr}\exp\{-\b{\bf H}\}=\frac 12
(Z_F^++Z_F^-+Z_B^+-Z_B^-),
\eeq
where $\b\equiv1/T$ is the inverse
temperature and the contributions on the right hand side are defined by
the formulae
\[Z_F^\pm=\mbox{Tr}\exp\{-\b{\bf H^\pm}\}=e^{\b
Mh/2}\det(I+J_\b);\]
\beq \label{(B.2)} Z_B^\pm=\mbox{Tr}[(-1)^{\bf
N}\exp\{-\b{\bf H^\pm}\}]e^{\b Mh/2}\det(I-J_\b), \eeq
where $J_\b$ is
a diagonal matrix
\beq \label{(B.3)}
J_\b=\mbox{diag}(e^{-\b\varepsilon(p)}),
\eeq
(we emphasize that $p\in X^+$ for $Z^+_{F,B}$
and $p\in X^-$ for $Z^-_{F,B}$). To obtain these representations
one should use the following relation
\beq
\label{(B.4)}
(-1)^{\bf N}|\phi,\pm\r=|-\phi,\pm\r,
\eeq
\[\exp\{-\b{\bf H^\pm}\}|\phi,\pm\r=e^{\b Mh/2}|J_\b\phi,\pm\r\equiv\]
\beq
\label{(B.5)}
\equiv e^{\b Mh/2}\exp\left\{\sl_{p\in X^\pm} a^+_pe^{-\b\varepsilon(p)}
\phi_p\right\}|0\r,
\eeq
and rewrite, for example,  $Z^+_{B}$ as a Gaussian integral
\beq
\label{(B.6)}
Z_B^\pm=e^{\b Mh/2}\int d\phi d\phi^*\exp\{\phi^*[I-J_\b]\phi\},
\eeq
(see \eqref{(A.4)}, \eqref{(A.5)}).

 Consider now the simplest equal-time temperature correlator,
\beq
\label{(B.7)}
G(m)\equiv\frac 1Z\mbox{Tr}\left[e^{\a{\bf Q}(m)}e^{-\b{\bf H}}\right],
\eeq
which can be represented in the following form \cite{cit}:
\beq
\label{(B.8)}
G(m)=\frac 1{2Z} (Z_F^+G_F^++Z_F^-G_F^-+Z_B^+G_B^+-Z_B^-G_B^-),
\eeq
where
\[Z_F^\pm G_F^\pm=\mbox{Tr}\left(e^{\a{\bf Q}(m)}e^{-\b{\bf
H^\pm}}\right),\]
\beq \label{(B.9)} Z_B^\pm
G_B^\pm=\mbox{Tr}\left(e^{\a{\bf Q}(m)}e^{-\b{\bf H^\pm}}(-1)^{\bf N}
\right).
\eeq
The contributions can be represented as determinants of
$M\times M$ matrices:
\[G_F^\pm=\det\left(I+(e^\a-1)Q^{(0)}(m)\Theta_F\right),\]
\beq
G_B^\pm=\det\left(I-(e^\a-1)Q^{(0)}(m)\Theta_B\right)
\label{(B.10)}
\eeq
where diagonal matrices of the Fermi and Bose weights
$\Theta_F$ and $\Theta_B$ are
\[\Theta_F=J_\b [I+J_\b ]^{-1}=\mbox{diag}\left[\Theta_F(p)\equiv
\frac{1}{1+e^{\b\varepsilon(p)}}\right],\]
\beq
\label{(B.11)}
\Theta_B=J_\b [I-J_\b ]^{-1}=\mbox{diag}\left[\Theta_B(p)\equiv
\frac{1}{e^{\b\varepsilon(p)}-1}\right],
\eeq
and the matrix $Q^{(0)}(m)$ with matrix elements $Q^{(0)}_{pq}(m)$was defined
in \eqref{(2.17)} ($p,q\in X^+$ for $G^+_{F,B}$
and $p,q\in X^-$ for $G^-_{F,B}$). To obtain these representations one
should rewrite, for example, the contribution $Z_B^-G_B^-$ as
\[Z_B^-G_B^-=e^{\b Mh/2}\int d\phi d\phi^*e^{\phi^*\phi}\l\phi^*,-|
e^{\a Q(m)}|-J_\b\phi,-\r,\]
use the expression \eqref{(2.18)} for the matrix element under the integral,
calculate the Gaussian integral, perform the similarity transformation
$Q(m)\rightarrow Q^{(0)}(m)$ (see \eqref{(2.16)}, \eqref{(2.17)}) and
extract the coefficient $\det (I-J_\b)$ from the determinant obtained.

Now we turn to the time-dependent correlation function of
the local spins; as usual
\[\sigma^\a_m(t)=e^{it{\bf H}}\sigma^\a_m e^{-it{\bf H}}.\]
Because of the translation invariance, the local spin correlators
$\sigma_{m_2}^\pm (t_2)$, $\sigma_{m_1}^\pm (t_1)$ depend only on
differences $m=m_2-m_1$, $t=t_2-t_1$:
\beq
\label{(B.12)}
G^{(ab)}(m,t)=\frac 1Z \mbox{Tr}(e^{it{\bf H}}\s^a_{m+1}e^{-it{\bf
H}}\s^b_1 e^{-\b{\bf H}}),\bl{1} a,b=\pm.
\eeq
Due to the property
\eqref{(2.21)} the correlator can be represented as a sum of four
contributions:
\beq \label{(B.13)} G^{(ab)}(m,t)=\frac 1{2Z}
\left(Z_F^+G_F^{(ab),+}+Z_F^-G_F^{(ab),-}
+Z_B^+G_B^{(ab),+}-Z_B^-G_B^{(ab),-}\right),
\eeq
where
\[Z_F^\pm G_F^{(ab),\pm}=\mbox{Tr}\left(e^{it{\bf H^\pm}}
\s^a_{m+1}e^{-it{\bf H^\mp}}\s^b_1e^{-\b{\bf H^\pm}}\right),\]
\beq
\label{(B.14)}
Z_B^\pm G_B^{(ab),\pm}=\mbox{Tr}\left(e^{it{\bf H^\pm}}\s^a_{m+1}
e^{-it{\bf H^\mp}}\s^b_1e^{-\b{\bf H^\pm}}(-1)^{\bf N}
\right).
\eeq

We begin with the correlator $G^{(-+)}(m,t)$. One can represent,
e.g., the contribution
\[Z_F^+G^{(-+)}=e^{\b Mh/2}\int d\phi d\phi^* d\psi d\psi^*
e^{\phi^*\phi-\psi^*\psi}\times\]
\[\times\l\phi^*J(0,t),+|\sigma^-_{m+1}|J(0,-t)\psi,-\r
\l\psi^*,-|\sigma_1^+|J_\b\phi ,+\r,\]
where we introduced a diagonal $M\times M$ matrix $J(m,t)$:
\beq
\label{(B.15)}
J(m,t)=\mbox{diag}\left(e^{-ipm+i\varepsilon(p)t}\right).
\eeq
Now we use an expression \eqref{(2.22)} for the matrix elements  and
perform the integration over $\psi,\psi^*$ and then over $\phi,\phi^*$:
\[Z_F^+G^{(-+)}=e^{\b Mh/2}\frac{\partial}{\partial\a}\int
d\phi d\phi^* d\psi d\psi^*
e^{\phi^*\phi-\psi^*\psi}\times\]
\[\left. e^{\phi^*L(m,t)\psi+\psi^*L(0)J_\b\phi +\a \phi^*RJ_\b\phi}\right|_
{\a =0}=\]
\[=\left.e^{\b Mh/2}\frac{\partial}{\partial\a}\det\left[I+
L(m,t)L(0)J_\b +\a RJ_\b\right]\right|_{\a=0},\]
where we defined the matrix
\[L(m,t)=J(0,t)L(m)J(0,-t)\]
(in  the matrix elements $L_{p,q}(m,t)$, $p\in X^+$ and $q\in X^-$),
and   also the matrix $R$ of rank 1 (all the columns of this
matrix are equal):
\[R_{p_1,p_2}=\frac 1Me^{-ip_1 m+i\varepsilon (p_1)t},\bl{1}p_1,p_2\in
X^+.\]
After the similarity transformation with the diagonal matrix
 $J(-\frac m2, -\frac t2)$ one gets \beq \label{(B.16)}
Z_F^+G_F^{(-+),+}=e^{\b Mh/2}\left.\pd{\a}\det\left[I+\tilde{L}J_\b+
\a\tilde{R}J_\b\right]\right|_{\a=0},
\eeq
where matrix elements of matrices $\tilde{L}$ and $\tilde{R}$ are
\beq
\label{(B.17)}
\tilde{L}_{p_1p_2}=e^{imp_1/2-it\varepsilon(p_1)/2}\sl_qL_{p_1q}(m,t)
L_{qp_2}(0)e^{-imp_2/2+it\varepsilon(p_2)/2},
\eeq
\beq
\label{(B.18)}
\tilde{R}_{p_1p_2}=\frac 1M e^{-imp_1/2+it\varepsilon(p_1)/2}
e^{-imp_2/2+it\varepsilon(p_2)/2},
\eeq
\[({\rm here}\bl{1} p_1,p_2\in X^+, q\in X^-).\]

To represent these matrices in a more convenient way we introduce
the following functions (we follow the paper \cite{cikt2}, changing
a bit some notations; recall that $\varepsilon (q)=h-\cos q$)
\beq
\label{(B.19)}
g(m,t)=\frac 1M\sl_qe^{imq+it\cos q};
\eeq
\beq
e(m,t,p)=\frac 1M\sl_q\frac{e^{imq+it\cos q}}{\tan\frac{q-p}{2}};
\label{(B.20)}
\eeq
\beq
d(m,t,p)=\frac 1M\sl_q\frac{e^{imq+it\cos q}-e^{imp+it\cos p}}
{\sin^2\frac{q-p}{2}};
\label{(B.21)}
\eeq
\beq
\label{(B.22)}
e_-(m,t,p)=e^{-i\frac{m}{2}p-i\frac{t}{2}\cos p};
\eeq
\beq
\label{(B.23)}
e_(m,t,p)=e_-(m,t,p)e(m,t,p).
\eeq
All these functions are of order $O(1)$ as $M\rightarrow\infty$.
It is also convenient to introduce four one-dimensional projectors
$\Pi^{++}$, $\Pi^{+-}$, $\Pi^{-+}$, $\Pi^{--}$ which are
$M\times M$ matrices with matrix elements
\beq
\label{(B.24)}
\Pi^{ab}_{p_1p_2}=e_a(m,t,p_1)e_b(m,t,p_2)\bl{1}((a,b)=+,-).
\eeq
Using identities
\beq
\label{(B.25)}
\sl_q\frac{1}{\sin^2\frac{q-p}{2}}=M^2\bl{1}(q\in X^-,p\in X^+);
\eeq
\beq
\label{(B.26)}
\cot(x-u)\cot(x-v)=\cot(u-v)[\cot(x-u)-\cot(x-v)]-1,
\eeq
one can represent the matrix $\tilde{L}$ in the following form
\beq
\label{(B.27)}
\tilde{L}=I+\frac 1M\left[S+i\Pi^{+-}-i\Pi^{-+}\right].
\eeq
The diagonal and nondiagonal matrix elements of the matrix $S$
are given by
\beq
\label{(B.28)}
S_{pp}=d(p)e^{-imp-it\cos p}
\eeq
\beq
\label{(B.29)}
S_{p_1p_2}=\frac{e_+(p_1)e_-(p_2)-e_-(p_1)e_+(p_2)}{\tan\frac{p_1-p_2}{2}}.
\eeq
The matrix $\tilde{R}$ is a projector,
\beq
\label{(B.30)}
\tilde{R}=\frac{e^{ith}}{M}\Pi^{--}
\eeq
(For  brevity,  the arguments $m$ and $t$ are omitted, being the same
for all the functions $d,e,e_{\pm}$).

Other contributions can be calculated analogously. We give only
the results:
\beq
\label{(B.31)}
G^{(-+),\pm}_F=e^{ith}\left.\pd{\a}\det\left[I+\frac 1M (S+i\Pi^{+-}-
i\Pi^{-+}+\a\Pi^{--})\Theta_F\right]\right|_{\a=0};
\eeq
\[G^{(-+),\pm}_B=e^{ith}\left.\pd{\a}\det\left[I+\frac 1M (S+i\Pi^{+-}-
i\Pi^{-+}+\a\Pi^{--})\Theta_B\right]\right|_{\a=0}.\]
Though formally both contributions $G_F^{(-+),\pm}$ (as $G_F^{(-+),\pm}$)
are written in the same form, really they are, of course, different.
It is necessary to take into account that
\[p,p_1 ,p_2\in X^+,\bl{1} q\in X^- \bl{1} {\rm for}\bl{1}
G_{F,B}^{(-+),+};\]
\[p,p_1 ,p_2\in X^-,\bl{1} q\in X^+ \bl{1} {\rm for}\bl{1}
G_{F,B}^{(-+),-};\]
in all the formulae \eqref{(B.19)}-\eqref{(B.24)}, \eqref{(B.28)} and \eqref{(B.29)} defining
the functions $g,e,d,e_\pm$ and the matrix elements of the matrices
$S$ and $\Pi^{ab}$.

Analogously one can calculate also the correlator $G^{(+-)}$. The
corresponding contributions are
\[G^{(+-),\pm}_F=e^{ith}\left[g(m,t)+\pd{\a}\right]\det\left[I+
\frac 1M (S+i\Pi^{+-}-i\Pi^{-+}-\vphantom{
-\a\{\Pi^{++}+ig\Pi^{+-}-ig\Pi^{-+}+g^2\Pi^{--}\})
\Theta_F}\right.\]
\beq
\label{(B.32)}
\left.\vphantom{\pd{\a}}\left.\vphantom{I+
\frac 1M (S+i\Pi^{+-}-i\Pi^{-+}-}
-\a\{\Pi^{++}+ig\Pi^{+-}-ig\Pi^{-+}+g^2\Pi^{--}\})
\Theta_F\right]\right|_{\a=0};
\eeq
\[G^{(+-),\pm}_B=e^{ith}\left[g(m,t)+\pd{\a}\right]\det\left[I+
\frac 1M (S+i\Pi^{+-}-i\Pi^{-+}-\vphantom{
-\a\{\Pi^{++}+ig\Pi^{+-}-ig\Pi^{-+}+g^2\Pi^{--}\})
\Theta_F}\right.\]
\[\left.\vphantom{\pd{\a}}\left.\vphantom{I+
\frac 1M (S+i\Pi^{+-}-i\Pi^{-+}-}
-\a\{\Pi^{++}+ig\Pi^{+-}-ig\Pi^{-+}+g^2\Pi^{--}\})
\Theta_B\right]\right|_{\a=0};\]
(one should take into account
once again the difference between the contributions
$G_{F,B}^{(+-),+}$ and $G_{F,B}^{(+-),-}$ explained above).

One should underline that the results obtained here coincide
with the representations obtained using another method
(for which rather complicated calculations are required)
in \cite{cikt2,cit}. Precisely our expression can be obtained
by choosing arbitrary constants $c_1=i, c_2=-i$ in the formulae
 \eqref{(A.1)}-\eqref{(A.10)} of the Appendix A of the paper \cite{iiks5}.


\begin{thebibliography}{30}
\bibitem{l1} A. Lenard {\it Momentum distribution in the ground
state of the one-dimensional sistem of impenetrable bosons.} - L. Math.
Phys. {\bf 5} (1964), 930-943.
\bibitem{l2} A. Lenard {\it One-dimensional impenetrable bosons
thermal equilibrium.} - J. Math. Phys. {\bf 7} (1966), 1268-1272.
\bibitem{ks} V.E. Korepin, N.A. Slavnov {\it The time-dependent
correlation functions of an impenetrable Bose gas as a Fredholm minor.} -
 Commun. Math Phys. {\bf  129} (1990), 103-113.
\bibitem{cikt1} F. Colomo, A.G. Izergin, V.E. Korepin and V. Tognetti,
{\it Correlators in the Heisenberg XX0 chain as Fredholm determinants.} -
Phys. Lett. {\bf A 169} (1992), 237-247.
\bibitem{cikt2} F. Colomo, A.G. Izergin, V.E. Korepin and V. Tognetti,
{\it Temperature correlation functions in the XX0 Heisenberg chain.} -
Teor. i Mat. Fiz. {\bf 94} (1993), 19-51.
\bibitem{k} V.E. Korepin, {\it Dual field formulation of quantum
integrable models.} - Commun. Math. Phys. {\bf 113} (1987), 177-190.
\bibitem{kks} T. Kojima, V.E. Korepin, and N.A. Slavnov, {\it Determinant
representation for dynamical correlation functions of the Quantum nonlinear
Schr\"odinger equation.} - Preprint hep-th/9611216, 26 Nov. 1996.
\bibitem{efik} F.H.L. Essler, H. Frahm, A.G. Izergin, and V.E. Korepin,
{\it Determinant
representation for correlation functions of spin 1/2 XXX and XXZ
Heisenberg magnets.} - Commun. Math. Phys. {\bf 174} (1995), 191-214.
\bibitem{jmms} A. Jimbo, T. Miwa, Y. Mori, and M. Sato,
{\it Density matrix of an impenetrable Bose gas and the fifth Painlev\'e
transcendent.} - Phys. {\bf D1} (1980), 80-158.
\bibitem{iik1} A.R. Its, A.G. Izergin, and V.E. Korepin,
{\it Correlation radius for one-dimensional impenetrable bosons.} -
Phys. Lett.  {\bf A 141} (1989), 121-125.
\bibitem{iiks1} A.R. Its, A.G. Izergin, V.E. Korepin, and N.A. Slavnov,
{\it Differential equations for quantum correlation functions.} -
Int. J. Phys. {\bf B4} (1990), 1003-1037.
\bibitem{iiks2} A.R. Its, A.G. Izergin, V.E. Korepin, and N.A. Slavnov,
{\it Temperature correlations of quantum spins.} - Phys. Rev. Lett.
{\bf 70} (1993), 1704-1706.
\bibitem{iiks3} A.G. Izergin, A.R. Its, V.E. Korepin, and N.A. Slavnov,
{\it Matrix Rieman-Hilbert problem and differential equations
for the correlation functions of the XX0 Heisenberg chain.} - Algebra i Analiz
{\bf 6} (1994), 138-151.
\bibitem{iik2} A.R. Its, A.G. Izergin, and V.E. Korepin,{\it Space correlations
on the one-dimensional impenetrable Bose gas at finite temperature.} -
Phys. {\bf D 53} (1991), 181-213.
\bibitem{iikv} A.R. Its, A.G. Izergin, V.E. Korepin, and G.G. Varguzin
{\it Large time and distance asymptotics of field
correlation functions of impenetrable bosons at finite temperature.} -
Phys. {\bf D 54} (1992), 351-395.
\bibitem{TMcC} C.A. Tracy, B.M. Mc Coy, {\it Neutron scattering and the
correlation function of the Ising model near $T_c$} - Phys. Rev. Lett.
{\bf 31} (1973), 1500-1504.
\bibitem{WMcC} T.T. Wu, B.M. Mc Coy, and C.A. Tracy,
{\it Spin-spin correlation function for the two-dimensional Ising model:
Exact theory in the scaling region.} - Phys. Rev. {\bf B 13} (1976), 316-374.
\bibitem{McCTW} B.M. Mc Coy, C.A. Tracy, and T.T. Wu,
{\it Painlev'e functions of the third kind.} -  J. Math. Phys. {\bf 18 }
(1977), 1058-1092.
\bibitem{bik} V.E. Korepin, A.G. Izergin, and N.M. Bogoliubov
{\it Quantum inverse scattering method and correlation functions.} -
Cambridge Monographs Math. Phys., Cambridge Univ. Press, Cambridge (1993).
\bibitem{iiks4} A.R. Its, A.G. Izergin, V.E. Korepin, and N.A. Slavnov,
 {\it The quantum correlation function as the tau-function of classical
differential equations.} - Springer Ser. Nonlinear Dyn., Springer, Berlin,
(1993), 407-417.
\bibitem{lsm} E. Lieb, T. Schultz, and D. Mattis {\it Two soluble models
of an antiferromagnetic chain.} - Ann. Phys. {\bf 16} (1961), 407-466.
\bibitem{McC} B.M. Mc Coy, {\it Spin correlation functions of the XY model.}
 - Phys. Rev. {\bf 173} (1968), 531-541.
\bibitem{McCBA} B.M. Mc Coy, E. Barouch, and D.B. Abraham, {\it Statistical
mechanics of the XY model. Time-dependent spin correlation functions.} -
Phys. Rev. {\bf A 4} (1971), 2331-2341.
\bibitem{McCPS} B.M. Mc Coy, J.H.H. Perk, R.E. Schrock {\it Correlation
functions of the transverse Ising chain at the critical field for large
temporaland spatial separations.} - Nucl. Phys. {\bf B 220 [FS 8]} (1983),
269-282.
\bibitem{ki} V.S. Kapitonov K.N. Ilinskii {\it Functional representation for the
correlators of the spin chains} - Zap. Nauch. Semin. POMI {\bf 224} (1995),
192-207.
\bibitem{cit} F. Colomo, A.G. Izergin, and V. Tognetti, {\it Correlation
functions in the XX0 Heisenberg chain and their relations with Frenkel
exciton spectra.} - J. Phys. {\bf A 30} (1997), 361-369.
\bibitem{iks} A.G. Izergin, N.A. Kitanine, and N.A. Slavnov, {\it On the
correlation functions of the XY model} - Zap. Nauch. Semin. POMI {\bf 224}
(1995), 178-191.
\bibitem{iiks5} A.G. Izergin, A.R. Its, V.E. Korepin, and N.A. Slavnov,
 {\it Integrable differential equations for the temperature correlation
functions of the XX0 Heisenberg chain} - Zap. Nauch. Semin. POMI
{\bf 205} (1993), 6-20.
\bibitem{ika} K.N. Ilinskii, and G.V. Kalinin, {\it Cyclic XY model and
exotic statistics in one dimension} - Phys. Rev. {\bf E 54} (1996),
R 1017-R 1020.
\bibitem{bie} N.V. Borisov, M.V. Ioffe, M.I. Eides, {\it Secondary
quantization of the field systems on a Grassman algebra} - Teor. i Mat.
Fiz. {\bf 29} (1976), 25-41.
%
\end{thebibliography}
\end{document}